\documentclass[10pt,
reprint,
 amsmath,amssymb,
 prl,
showpacs,
floatfix,
lengthcheck,%
]{revtex4-1}
\usepackage{graphicx}
\usepackage{epstopdf} 
\usepackage{dcolumn}
\usepackage{bm}
\usepackage{natbib}
\usepackage[normalem]{ulem}

\begin{document}

\title{Carbon nanotube plectonemes: Loops of twisted helices}
\author{Alireza Shahabi and Moneesh Upmanyu}
\email{mupmanyu@neu.edu}
\affiliation{Group for Simulation and Theory of Atomic-Scale Material Phenomena ({\it st}{\rm AMP}), Department of Mechanical and Industry Engineering, Northeastern University, Boston, Massachusetts 02115, USA.}

\begin{abstract}
The relaxation of twist in elastic filaments often drives conformational changes. 
We explore this paradigm using all-atom computations and report the formation of novel supercoiled shapes in individual carbon nanotubes (CNTs). Decreasing the end distance of torsionally constrained CNTs leads to spontaneous nucleation and growth of a nanotube plectoneme. We develop a stability diagram and comparisons with theoretical frameworks reveal the importance of non-local van Der Waals interactions. In some cases, they stabilize the supercoiling to an extent that its tip locally kinks and then irreversibly reconstructs into a disordered yet strengthened structure that involves $sp^3$ bonding. The ability to engineer supercoiled conformations of CNTs and related nanoscale filaments opens the possibility of a unique set of tunable functional properties at the nanoscale. 
\end{abstract}

\maketitle

The interplay between structure, morphology and mechanics that sets the conformations of elastic filaments is important in both nature and technology. As an example, twist storing soft filaments such as DNA and filamentous protein assemblies can lower the elastic energy associated with overtwist by supercoiling~\cite{biofil:StrickBensimon:1998, biofil:MorozNelson:1998}, and the conformational control is crucial in both health and disease. The morphology of crystalline nanoscale filaments - nanotubes, nanowires and nanoribbons - is equally important as it shapes their properties, and is therefore crucial for next-generation applications that rely on these building blocks as active elements. As the lengths of the as-synthesized nanofilaments approach macroscale dimensions and become comparable to their persistence lengths for bending and twisting, their  conformations are expected to have strong parallels with those of their semi-flexible counterparts. Indeed, coiled and double-helical shapes have been recently reported in ultralong fibrous assemblies of nanowires and nanotubes~\cite{nw:WangChen:2011, *cnt:ZhengZhu:2004, *nw:ShangCao:2013, *fil:YangKotov:2011}, yet control over their conformation remains limited and serves as the main motivation for this study.
 


In this Letter, we use single-walled CNTs as model systems to explore supercoiled soft conformations in individual nanoscale crystalline filaments via torsional constraints. All-atom molecular dynamics (MD) computations of finite length CNTs with a quenched twist density $\phi$ are employed to identify energy minimizing conformations, for prescribed end distances $z$ smaller than their contour length $L$, $\lambda=z/L<1$, and a prescribed net rotation $\Phi=\phi L$ across the ends. The choice is dictated by the fact that, unlike end couple (moment $M$ and axial tension $T$), the end distance and rotation are easily accessible in experiments on individual nanoscale filaments~\cite{nt:FenimoreZetl:2003, *nt:Bachtold:2004, *nt:PapadakisSuperfine:2004, *asit:CohenJoselevich:2006}. For generalized filaments so constrained, the combination of fixed end displacement and end rotation results in totally rigid loading, and the partitioning of the elastic energy density preserves a global topological invariant associated with the supercoiled geometry, the Linking number $Lk$, given by the well-known C{\u a}lug{\u a}reanu-White-Fuller theorem, $Lk = Tw + Wr$. The Twist $Tw$ is the rotation of a material frame about the local tangent and the Writhe $Wr$ is a measure of the bent geometry of the filament axis~\cite{elastica:Calugareanu:1961, *topol:White:1969, *elastica:Fuller:1971}. 

We study this interplay in three achiral CNTs: (3,3), (6,6) and (8,8) armchair nanotubes with diameters $D=0.434$, $0.826$ and $1.094$\,nm, respectively. In each case, the length of the nanotube is fixed, $L=110.4$\,nm. The CNTs are torsionally constrained in the range $Lk=1.25-8.75$; the corresponding to twist densities are in the range $\phi=4.1-28.5^\circ$\,nm$^{-1}$.
The initially twisted configuration is generated by uniformly rotating the two ends by an angle $\Phi$. The ends are clamped to fix $Lk=\Phi/2\pi$, also equal to the $Tw$ in the initially straight CNT. The formation and stability of the conformations is then explored by decreasing the end distance at a rate $\delta z/\delta t \approx4$\,nm/ns within room temperature canonical MD simulations performed using an AIREBO carbon-carbon potential~\cite{intpot:Brenner:2002, md:Plimpton:1995, suppdocs}. 

\begin{figure*}[htp]
\includegraphics[width=1.9\columnwidth]{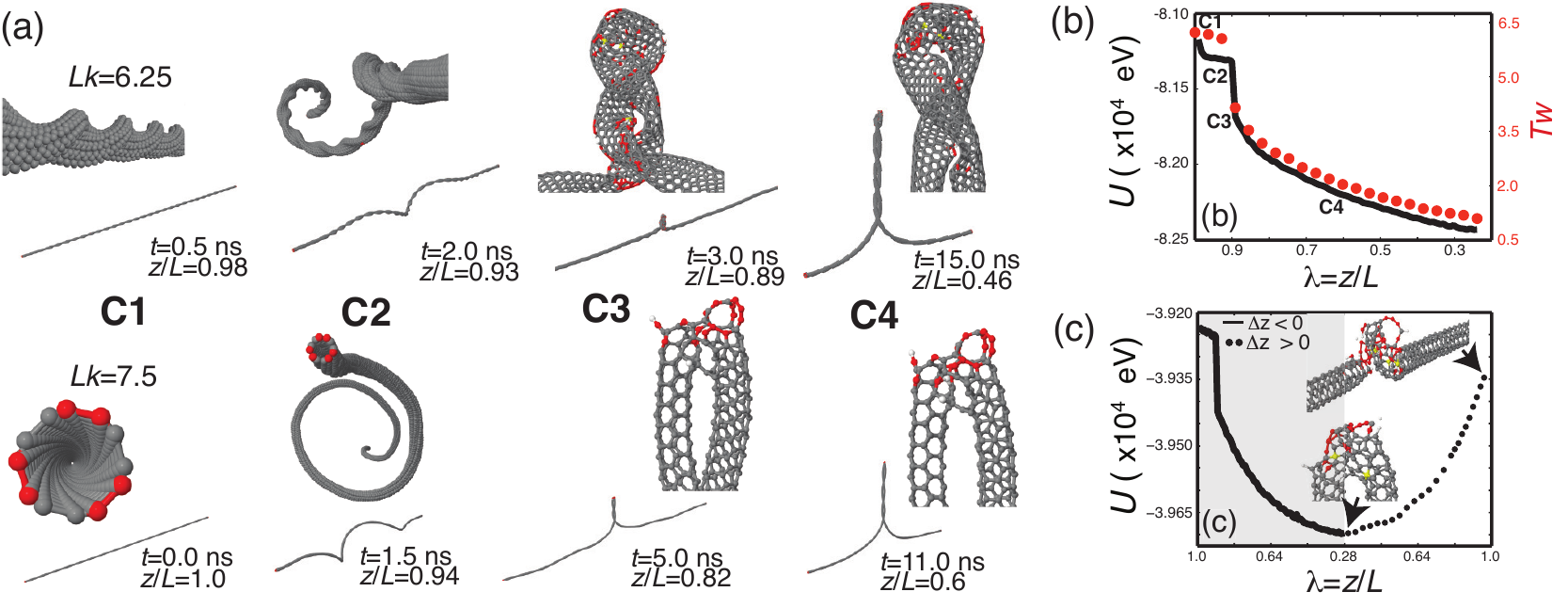}
 \caption{(color online) (a) (top row) Sequence of configurations (labeled C1-C4) observed in MD simulations of a (6,6) CNT with Linking number $Lk=6.25$ and decreasing end distance $\lambda=z/L$. 
 Expanded views of the configurations are shown alongside;  they are rotated to depict the details. Here and elsewhere, the atoms are colored based on the number of neighbors within a cut-off distance of 1.5\,{\rm \AA}: white-1, red-2, grey-3 and yellow-4. (bottom row) Same as above but for a (3,3) CNT with $Lk=7.5$. (b) Variation in total interaction energy $U(\lambda)$ and the twist $Tw(\lambda)$ for the (6,6) CNT in (a). The locations of the configurations C1-C4 in (a) are maked. (c) $U(\lambda)$ for a (3,3) CNT with $Lk=7.5$, for decreasing (solid curve) and increasing (dotted curve) end distance $z$. (inset) Magnified views of kinked tip of the plectoneme at $\lambda=0.28$, and at $\lambda=1$ following the re-extension~\cite{suppdocs}.  
\label{fig:figure1}
 }
\end{figure*}

Figure~\ref{fig:figure1}a depicts four commonly observed conformations, henceforth labeled C1-C4, within a (6,6) CNT with $Lk=6.25$. The relatively large $Lk$ induces torsional buckling (top row, C1) as the initial twist density is high~\cite{cnt:WangVaradan:2007}. As expected, this is more frequent at large diameters. It is stable for $\lambda\approx1$, and further decreasing $\lambda$  lowers the tension induced by the twist and it destabilizes into a helix (C2, Fig.~\ref{fig:figure1}a). The curvature of the CNT axis increases at the expense of the twist density and the extent of buckling therefore decreases (inset). Thereafter, the work done in decreasing $\lambda$ increases the tortuosity of the CNT axis. 
Below another critical $\lambda^\ast$ the helix develops a local instability that leads to the nucleation of a supercoiled phase, a nanotube plectoneme (C3, Fig.~\ref{fig:figure1}a). The transition is spontaneous as it is immediately stabilized by non-local van der Waals interactions between the helically coiled CNT surfaces. The torsional buckling is reduced in extent and is mostly confined to 
the plectoneme; the remainder of the CNT is almost straight and twisted. 

The CNT plectoneme is unlike the supercoils observed in soft filaments~\cite{biofil:BolesCozzarelli:1990, *biofil:BancaudPrunell:2006, *biofil:GhatakMahadevan:2005}. 
The looped terminal end  is relatively smaller in extent and tightly curved over a radius of the order of the CNT diameter, i.e. it resembles a kink. It is not an artifact of the large strain rates; an order of magnitude decrease in the strain rate has no effect. Smaller twist densities $2\pi Lk/L$ result in large radii helices that again collapse into kinked plectonemes~\cite{suppdocs}, suggesting that the release of the torsional energy is size constrained such that the plectoneme tip cannot form a much larger sized loop. Interestingly, the kink is structurally defective (inset, Fig.~\ref{fig:figure1}c). 
The changes in the bonding network include the formation of defect clusters that absorb the large curvature~\cite{suppdocs}. In addition to several dangling bonds (white) and non-linearly strained atoms (red), we observe the formation of several clusters each containing an $sp^3$ bonded carbon atom at its core (yellow). In this regard, the kink formation is unlike prior reports of kink formation in bent and buckled CNTs~\cite{nt:Yakobson:1996, fil:CohenMahadevan:2003} as it is stabilized by the non-local van der Waals interactions. Further decreasing $\lambda$ results in the growth of the plectoneme. The defective structure of the kink remains mostly unchanged, indicating that the modified bond network is stable. Away from the tip, the two CNT segments form a double helix with a radius of the order of the CNT diameter $r\approx D/2$ and a much larger pitch length, $p\approx20r$.

Figure~\ref{fig:figure1}a (bottom row) shows the response of a (3,3) CNT with $Lk=7.5$. Although these small diameter CNTs are heavily strained and are unstable during catalyzed growth, they do not exhibit torsional buckling and represent model systems for studying the interplay between twist and bending at smaller lengths accessible to the simulations.  As before, the initially straight and twisted conformation destabilizes into a helix (C2, $\lambda=0.94$). The nucleation of the plectoneme is relatively delayed (C3, $\lambda=0.82$). The looped end is small and defective with several vacancy clusters that lead to partial fracture at the kink, and dangling bonds that reconstruct into short bridges~\cite{suppdocs}. The supercoiled region progressively lengthens with decreasing $\lambda$ and a pitch length $p\approx25r$ (C4).

Fig.~\ref{fig:figure1}b shows the combined plot of the net interaction energy $U$ and Twist $Tw$ as a function of $\lambda$ for the (6,6) CNT (Fig.~\ref{fig:figure1}a). The energy evolution $U(\lambda)$ shows four distinct regimes: i) a rapid decrease as the twisted CNT transitions into a helix, ii) a slower linear decrease as the helix increases its radius and then localizes, iii) an abrupt and large decrease at $\lambda\approx0.9$ consistent with the spontaneous nucleation of the plectoneme, iv) and a slow non-linear decrease as the plectoneme grows. The stored twist $Tw$ is extracted dynamically~\cite{suppdocs} and exhibits similar trends.
\begin{figure}[t]
\includegraphics[width=\columnwidth]{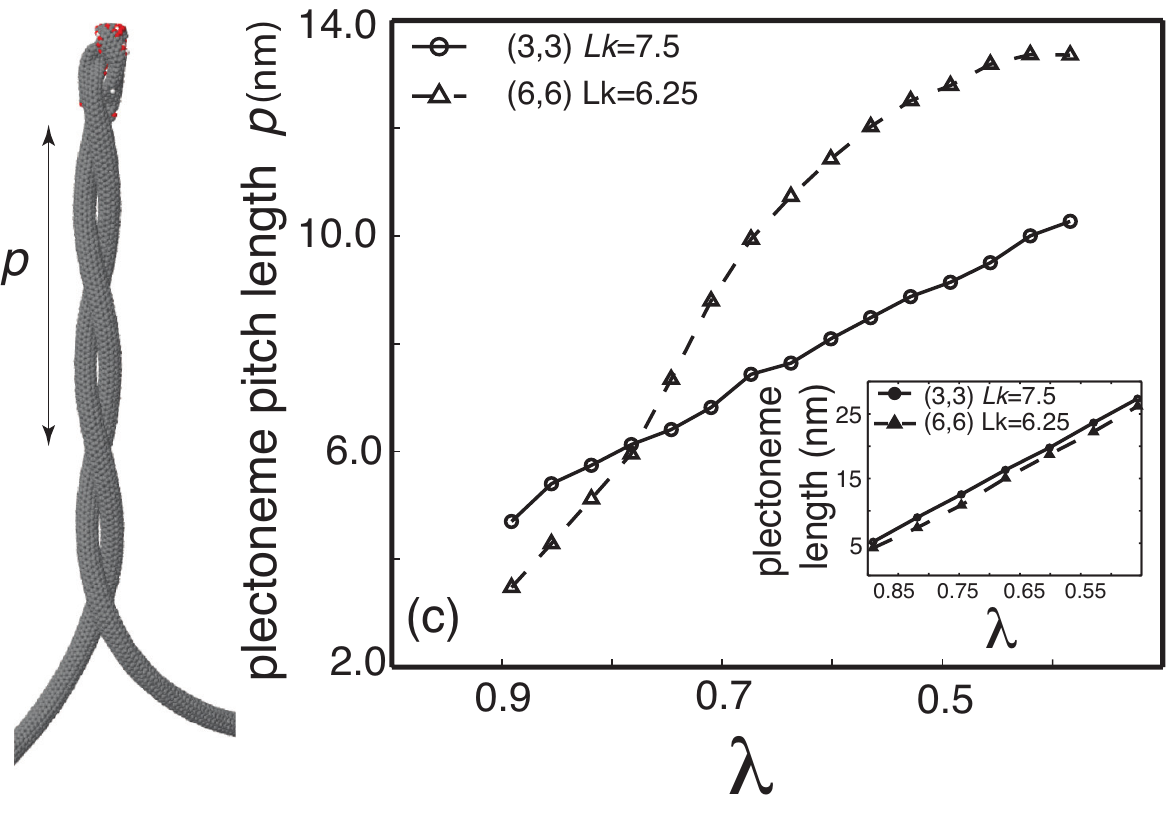}
 \caption{The evolution of the pitch length $p$ of the plectoneme with decreasing $\lambda$ in supercoiled CNTs shown in Fig.~\ref{fig:figure1}. (inset)  Variation in the tip-to-base plectoneme length with $\lambda$. \label{fig:figure2}}
\end{figure}

The initial linear decrease is due to the increasing Writhe as the helical phase forms and grows. The nucleation of the plectoneme leads to a larger decrease at $\lambda\approx0.9$, followed by a gradual non-linear decrease that closely follows the change in energy as the plectoneme grows in length. The response of the (3,3) CNT is quite similar (Fig.~\ref{fig:figure1}c) with some minor variations that arise primarily because of the absence of buckling. Both $U$ and $Tw$ decrease smoothly during the helix formation. The transition to the plectoneme occurs at $\lambda\approx0.9$, yet the nucleated plectoneme is longer in length and the decrease in energy is larger as the resultant bond network consists of several $sp^3$ bonds. We have tested the stability of the kinked plectoneme by re-extending the CNT. The energy increases non-linearly without any discontinuities as the plectoneme unwinds within an otherwise straight CNT. At $\lambda=1$, the defective kink persists and both $U$ and $Tw$ (not shown) do not recover completely. Interestingly, $U$ is lowered following re-extension, a clear indication that the defective region is stabilized and strengthened by the bond network.

Figure~\ref{fig:figure2} shows the evolution of the plectoneme post-nucleation. The pitch length $p(\lambda)$ is not constant, rather it increases for both (3,3) and (6,6) CNT. The increase is linear, $p\sim\lambda$, although the response for the (6,6) CNT is more rapid as it is buckled. Evidently, the plectoneme becomes less intertwined as it grows, and this indicates that i) the CNTs slide with respect to each other, and ii) there is no twist stored within the plectoneme~\cite{suppdocs}. 

A simple analysis yields insight into the stability of the plectoneme; in the absence of twist and stretching strains within the plectoneme, the energetics reduces to a competition between bending and interaction energies and the work done by the external couple. The bending energy scales as $U_b(p)\sim (D^2\kappa_b/p^4) l$ while the interaction energy scales as $U_i(p)\approx (a u_0) l + \mathcal{O}(p^{-2})$, where $\kappa_b$ is the bending rigidity, $u_0$ is the interaction energy per unit area between graphene elements and $a$ is the width of the approximately straight interaction area for the large $p/r$ ratios observed here~\cite{cntr:LiangUpmanyu:2005b, biofil:PokroyAizenberg:2009}. The  base-to-tip plectoneme length also grows linearly, $l\sim\lambda$ (inset). Ignoring changes in the kink structure and the curvature of plectoneme base that connects to the ends, the energy stored in the double helical region evolves as $U\sim D^2\kappa_b/\lambda^3 - au_0 \lambda$. For CNTs, $k_b/u_0>1$ and $a\sim r$~\cite{cntr:Girifalco:2000}. Clearly, the interaction energy is insufficient and the double-helical segment is stabilized by the external work done by the tension $T\lambda$ and the release of the torsional energy $M\phi$. 


\begin{figure}[bh]
\includegraphics[width=\columnwidth]{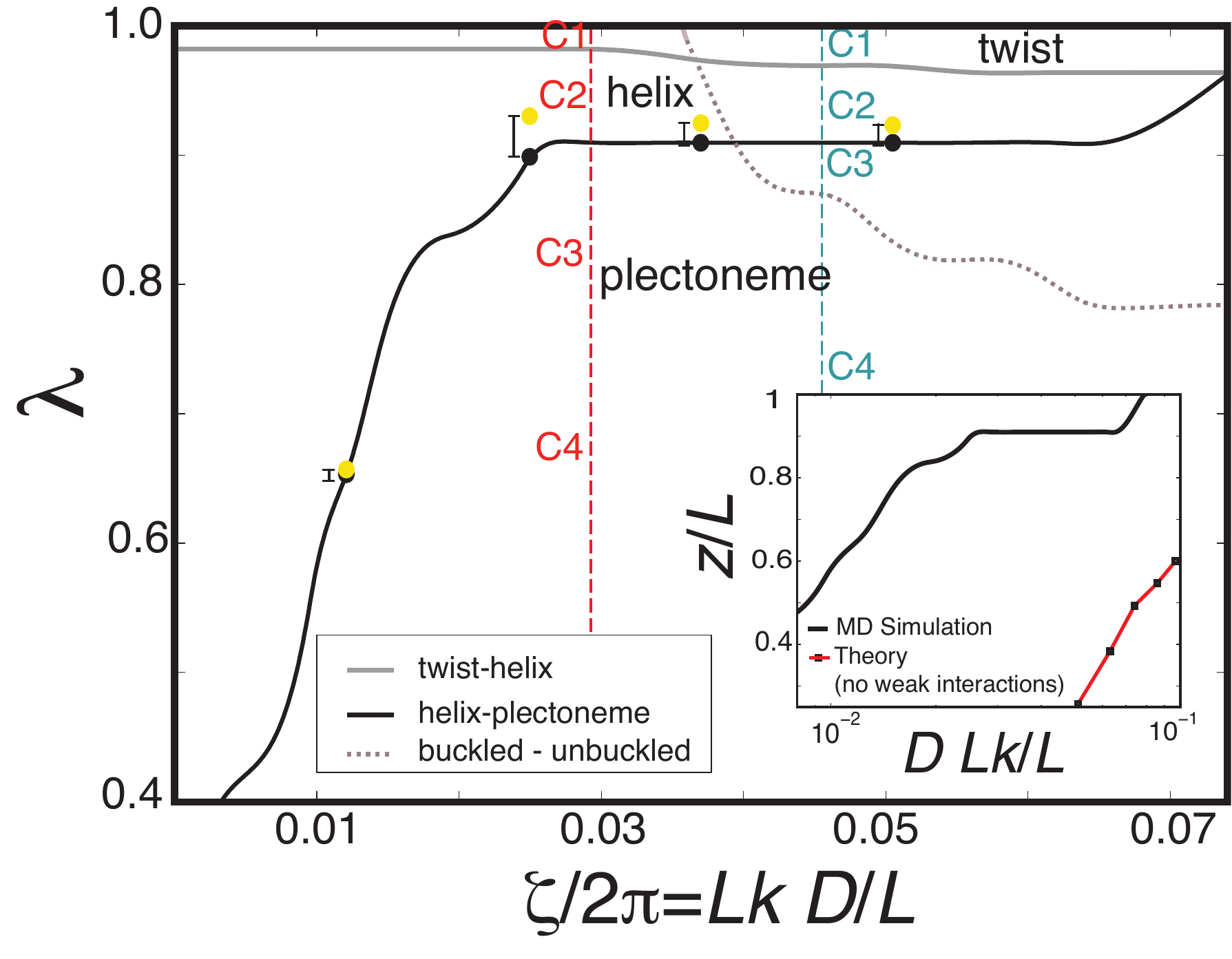}
 \caption{(color online) Dimensionless conformational diagram for CNTs with varying diameters $D$ and Linking numbers $Lk$. The solid curves represent co-existence lines for the three phases observed in the simulations: twist, helix, and plectoneme. Torsionally buckled CNTs are stable above the dotted curve (high $\zeta$ and $\lambda)$. The red and blue vertical lines correspond to the achiral CNTs shown in Fig.~\ref{fig:figure1}: (3,3) with $Lk=7.5$, and (6,6) with $Lk=6.25$, respectively. The points at which configurations C1-C4 reported in Fig.~\ref{fig:figure1} first appear are also indicated. Yellow circles represent the helix-plectoneme transition of (6,6) CNTs with longer lengths, $L=166$\,nm. (inset) The helix-plectoneme transition observed in the simulations and that predicted by theory for non-interacting elastic filaments~\cite{elastica:Coyne:1990}; see text for details. \label{fig:figure3}}
\end{figure}
In order to develop a more complete picture of the stability of and transitions in the conformations, we have studied the response of several achiral CNTs with varying $Lk$. The results are plotted as a dimensionless conformational phase diagram, $\lambda$ versus the size-weighted link density, $\zeta/2\pi=Lk (D/L)$ (Fig.~\ref{fig:figure3}). The two solid boundary lines define regions of stability of the three conformations - twist, helix, and plectoneme. The region enclosed by the dashed line (top right) indicates completely or partially buckled conformations at larger CNT diameters and $Lk$.  The transition is length independent; we have performed four additional simulations of (6,6) CNTs with length of $166$\,nm ($=1.5L$). The critical point for plectoneme formation is almost unchanged (yellow circles). 

\begin{figure}[tb]
\includegraphics[width=0.95\columnwidth]{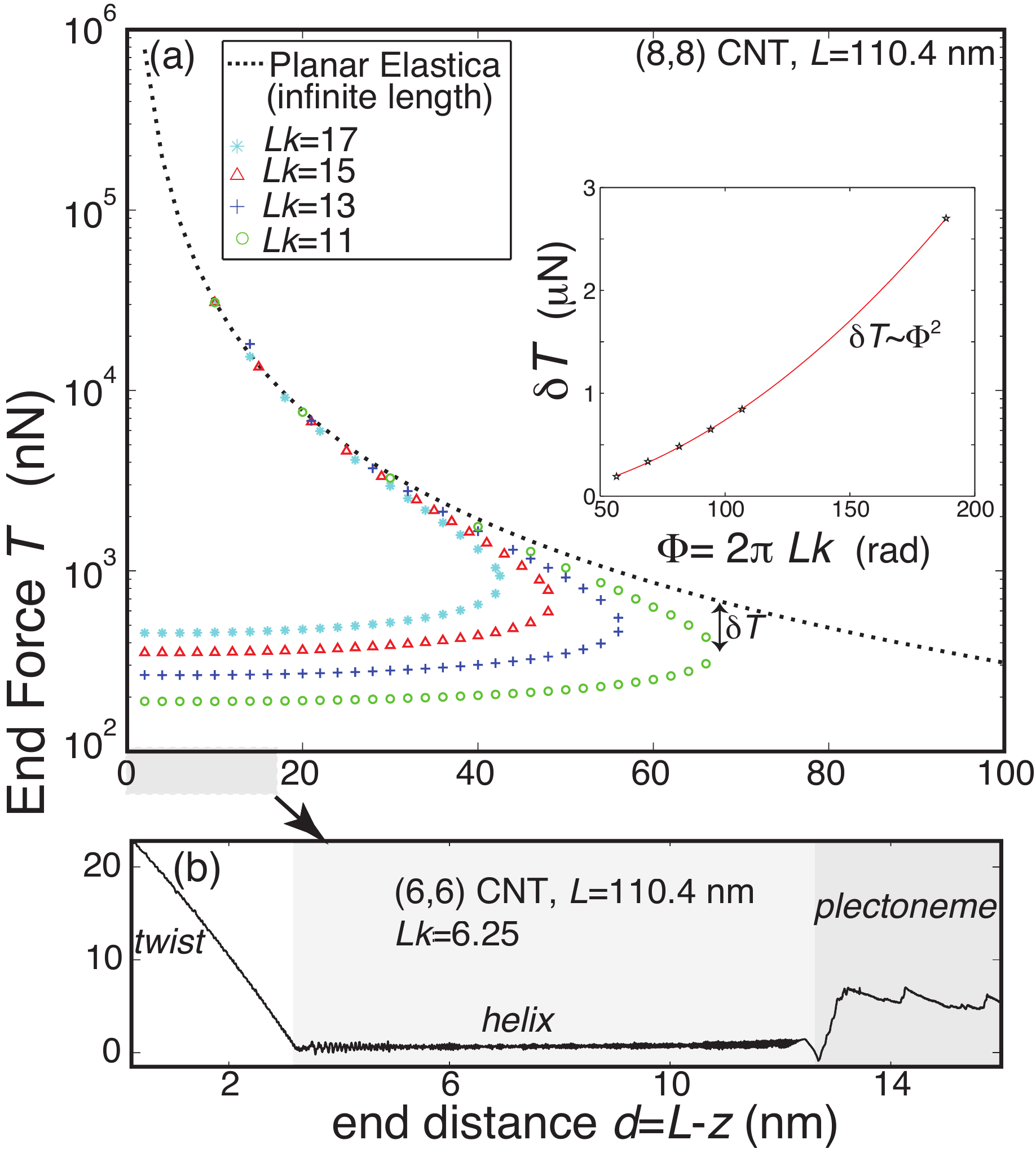}
 \caption{Force-displacement curves based on Eq.~\ref{eq:CoynesResult} for the $(8,8)$ supercoiled CNT and varying $Lk$~\cite{elastica:Coyne:1990}. The lower arm in each curve is the helical phase and the upper arm is the localizing instability that leads to plectoneme formation. The nose of the curve is the critical point $(T^\ast, d^\ast)$ for the helix-plectoneme transition. The dashed line is the planar-looped elastica solution for an infinite elastic rod. (inset) Correction to the planar-looped elastica solution $\delta T$ for the critical point as a function of $Lk$. (b) The evolution of the end tension $T(d)$ extracted in the simulations.
 \label{fig:figure4}}
\end{figure}

The twist-helix transition occurs for $\lambda\approx1$ and is independent of the link density $\zeta$. Evidently, the twist is stabilized by large axial tension which decreases linearly with $\lambda$, as confirmed later (Fig.~\ref{fig:figure4}b). Below a critical tension, the helix is energetically preferred.  At large $Lk$, the transition is mediated by buckling which lowers the critical $\lambda$. 

The helix-plectoneme transition varies dramatically. The critical point $\lambda^\ast$ increases with the critical link density $\zeta^\ast$ and then saturates as we approach the buckling threshold. The trend is as expected since the localization of the helix that results in the plectoneme is facilitated by the initial twist. To develop a more detailed understanding, we follow the energetic analysis performed by Coyne~\cite{elastica:Coyne:1990} on a finite-length elastica subject to an end couple ($M$, $T$). Minimization of the energies stored in bending and twisting and the work done by the end tension (for fixed $Lk$ there is no work done by the moment) as a function of the end displacement, the critical point $(T^\ast, \beta^\ast)$ at which the helix transitions into a localized loop is~\cite{elastica:Coyne:1990} 
\begin{align}
\label{eq:CoynesResult}
T = \frac{\kappa_t^2}{4\kappa_b L^2} \frac{\left(\phi L - 4 \sin^{-1}\beta\right)^2}{1-\beta^2},
\end{align}
where $\kappa_t$  is the torsional rigidity and $\beta$ is a dimensionless end displacement defined as $\beta=d\sqrt{T/16\kappa_b}$. Figure~\ref{fig:figure4}a shows the theoretically predicted $T(d, Lk)$ for an $(8,8)$ CNT. For fixed $Lk$, the variation $T$ vs. $d$ is bistable and the nose of each curve where the helical shape transitions to the localizing loop (precursor to the plectoneme) is the critical point~\cite{elastica:ThompsonChampneysI:1996, *elastica:HeijdenThompson:2000}. 
For comparison, we have also plotted the solution for an infinitely long planar looped elastica, $T=16\kappa_b / d^2$, which overestimates the critical tension; the difference $\delta T$ increases quadratically with $Lk$ (inset). Correcting the planar solution accordingly yields a general solution for the critical tension $T^\ast(d^\ast, \zeta^\ast)$~\cite{suppdocs}. The critical points $\lambda^\ast=1-(d^\ast/L)$ and $\zeta^\ast$ are plotted in the phase diagram (inset, Fig.~\ref{fig:figure3}). The trends are similar but do not agree quantitatively.  
Refinements such as the effect of clamped edges and more detailed analyses that absorb the shape of the plectoneme~\cite{elastica:GoyalLee:2005, *elastica:Purohit:2008} can mitigate the discrepancy,  yet the fact that the theory overestimates the critical end distance and link density suggests a non-trivial effect of long-range interactions that is ignored in the classical elastica solutions and enhance the plectoneme formation. As an analogous example, recent studies on DNA supercoiling that consider electrostatic interactions are in agreement with experiments~\cite{biofil:ClauvelinNeukirch:2009}, and similar corrections to classical elastic analysis are necessary for a quantitative understanding of the plectoneme formations in CNTs, and more general nanoscale filaments. 

In order to quantify this effect, we extract the end tension dynamically for the (6,6) CNT with $Lk=6.25$ (Fig.~\ref{fig:figure4}b). The tension, of the order of tens of nN, is smaller compared to the plots shown in Fig.~\ref{fig:figure4}a as the $Lk$ is itself smaller. The twist-helix transition follows the linear decrease in the tension which then stabilizes to a few nN. The localized instability results in an abrupt decrease in the tension,  large enough to change its sign. The transition occurs at almost constant $\lambda$ and the growth of the double-helix is reined in by the abrupt increase in the tension that eventually exceeds that in the helical phase. Thereafter, the tension varies in a serrated  fashion for every new loop absorbed by the plectoneme. These transitions are gradual, and each cycle involves a balance between long-range interactions that  facilitate loop formation followed by recovery of the tension as the  plectoneme grows and adjusts its pitch length.

In conclusion, we present a novel paradigm for engineering plectonemic phases in individual CNTs. Our results have implications for supercoiled conformations in nanoscale crystalline filaments such as nanowires and nanoribbons, where the combination of crystalline order and long-range attractive interactions is equally important. The necessary axial and torsional manipulation can be realized via exploiting the coupling that naturally exists in chiral nanotubes and their ropes~\cite{asit:LiangUpmanyu:2006, *asit:LiangUpmanyuMahajan:2008, *cntr:LiangUpmanyu:2005b}, electromechanical actuation based on paddle functionalization~\cite{nt:FenimoreZetl:2003, *nt:Bachtold:2004, *asit:CohenJoselevich:2006}, atomic force microscope tips~\cite{nt:WilliamsSuperfine:2002, *nt:PapadakisSuperfine:2004}, chemical and photonic manipulation~\cite{cnt:LimaBaughman:2012}, and transfer onto appropriately pre-deformed substrates~\cite{nr:SunRogers:2006, *nw:XuZhu:2010, *nt:SomuUpmanyu:2010, *cnt:HahmWangUpmanyuJung:2012}. Although details of the plectoneme structure await experimental confirmation, their controlled nucleation and growth opens up the possibility of a unique set of multifunctional properties - electronic, mechanical, thermal, optical - that remain to be harnessed in nanofilament-based device platforms. The quantitative understanding developed in this study allows control over the extent of supercoiling that translates to {\it on-demand} and reversible modulation of these properties, a key enabler for robust and highly non-linear transport, actuation, switching and energy storage at the nanoscale.

{\bf Acknowledgements}: The authors acknowledge support from National Science Foundation DMR CMMT (\#1106214, AS and MU) and DMREF CHE (\#14348424, MU) Programs. The computations were performed on {\it st}AMP and Discovery (MGHPCC) supercomputing resources at Northeastern University.
%

\end{document}